\documentclass[9pt]{article}

\usepackage{lineno}

\usepackage[utf8]{inputenc}
\usepackage[english]{babel}
\usepackage{amssymb}
\usepackage{amsmath}

\usepackage[margin=0.75in, top=0.75in]{geometry}

\usepackage{framed}
\usepackage{authblk}

\usepackage{enumitem}
\newlist{todolist}{itemize}{2}
\setlist[todolist]{label=$\square$}

\usepackage[
backend=biber,
style=alphabetic,
citestyle=authoryear
]{biblatex}
\usepackage[colorlinks,citecolor=red,urlcolor=blue,bookmarks=false,hypertexnames=true]{hyperref}

\usepackage{graphicx}
\graphicspath{ {./images/} }
\usepackage{subcaption}

\title{Efficient Computation for Centered Linear Regression with Sparse Inputs}
\author{Jeffrey Wong\\
  Experimentation Platform\\
  Netflix, Inc.
}
\date{\today}

\begin{document}

\maketitle

\begin{abstract}
    Regression with sparse inputs is a common theme for large scale models. Optimizing the underlying linear algebra for sparse inputs allows such models to be estimated faster. At the same time, centering the inputs has benefits in improving the interpretation and convergence of the model. However, centering the data naturally makes sparse data become dense, limiting opportunities for optimization. We propose an efficient  strategy that estimates centered regression while taking advantage of sparse structure in data, improving computational performance and decreasing the memory footprint of the estimator.
\end{abstract}

\section{Introduction}

Centering regression inputs is an operation done when estimating linear models, such as Ordinary Least Squares (OLS) that can improve interpretation and convergence. Given the model $y = \beta_0 + X\beta_1 + \epsilon$ with $\epsilon \sim N(0, V)$, 
centering removes the average of $X$ from each feature vector in the dataset so that the centered feature represents the deviation from the average. This allows the parameter, $\hat{\beta}_0$, to represent the mean of $y$ for the average observation. The coefficients, $\hat{\beta}_1$, then represent offsets from the average observation (\cite{aiken1991multiple}). In randomized and controlled trials, we often seek the average treatment effect, which is the average deviation between observations where treatment was applied and observations where treatment was not applied. Without centering, $\hat{\beta}_0$ represents the mean of $y$ for an arbitrary baseline group, and $\hat{\beta}_1$ would be the offset from that arbitrary baseline.

When the model uses features that are transformations of $X$, for example, $X^2$, centering the regression makes it easier to estimate the distribution of $\hat{\beta} = (\hat{\beta}_0, \hat{\beta}_1)$. Removing the centers of the features shrinks the covariances between the features and decreases the condition number. As the covariances shrink, the covariance matrix behaves more like a diagonal matrix, which is easy to invert and solve. 

Optimizing regression models for sparse inputs is an important component of performance (\cite{wong2019efficient}), especially when using one hot encoded categorical variables, however it can conflict with centered regression. Given a model matrix with arbitrary features, $M$, and a response vector, $y$, we wish to estimate OLS with centered inputs. The matrix $M$ can be stored as a dense matrix, $M_D$, or as a sparse matrix, $M_S$. If $M$ is stored as a dense matrix, centering the inputs, then estimating OLS incurs little overhead. However, if $M$ is stored as a sparse matrix, centering $M_S$ results in a dense matrix, limiting sparse optimizations that take advantage of the structure of $M_S$. In this paper we show how to expand OLS so that we take advantage of the structure of $M_S$ as much as possible. We show this expansion for three common cases, and then describe linear algebra optimizations to create an efficient solver.

\section{Expanding OLS}

Let $M_S \in \mathbb{R}^{n \times p}$ be a model matrix, $\mu^T \in \mathbb{R}^{1 \times p}$ be a row vector for the $p$ column means of $M_S$, and $1_{n}$ be a length $n$ column vector of all ones. Then the centered model matrix is $\tilde{M} = (M_S - 1_n\mu^T)$. Estimating OLS with $\tilde{M}$ and $y$ yields the estimator 

$$\hat{\beta}_{\text{transformed}} = (\tilde{M}^T \tilde{M})^{-1} \tilde{M}^T y.$$

This can be done by materializing $\tilde{M}$ and passing it as input to a standard OLS solver, such as StatsModels (\cite{seabold2010statsmodels}). However, materializing $\tilde{M}$ transforms the inputs from sparse to dense. Instead, we use a strategy that splits operations into sparse optimized operations, and dense operations that are easy to compute. This section describes the linear algebra operations for three different use cases, then section 4 describes the optimizations. To better utilize the sparsity in $M_S$ we write the first expansion as

\begin{align*}
  \hat{\beta}_{\text{transformed}} &= \Big((M_S - 1_n\mu^T)^T (M_S - 1_n\mu^T)\Big)^{-1} (M_S - 1_n\mu^T)^T y \\
  &= \Big(M_S^T M_S - M_S^T 1_n\mu^T - \mu 1_n^T M_S + \mu 1_n^T 1_n \mu^T\Big)^{-1} (M_S^T y - \mu1_n^Ty)  
\end{align*}

where $M_S^T M_S$ is a sparse optimized matrix multiplication, and $- M_S^T 1_n\mu^T - \mu 1_n^T M_S + \mu 1_n^T 1_n \mu^T$ is an easy to compute dense operation, discussed in section 4.

\subsection{Weighted OLS}

Suppose the regression is weighted by the vector $w$. Then centering refers to removing the weighted mean $1_n \mu_w^T$ where $\mu_w^T$ is the weighted column means of $M_S$ weighted by $w$. The weighted OLS problem becomes

\begin{align*}
  \hat{\beta}_{\text{transformed}} &= \Big((M_S - 1_n\mu_w^T)^T W (M_S - 1_n\mu_w^T)\Big)^{-1} (M_S - 1_n\mu_w^T)^T W y \\
  &= \Big(M_S^T W M_S - M_S^T W 1_n\mu_w^T - \mu_w 1_n^T W M_S + \mu_w 1_n^T W 1_n \mu_w^T\Big)^{-1} (M_S^T W y - \mu_w 1_n^T W y).
\end{align*}

\subsection{Scaling and Centering}

In addition to centering, we may also wish to scale $M_S$ so that each column has weighted variance 1. For instance, the lasso (\cite{tibshirani1996regression}) shrinks coefficients to zero and assumes data has been centered and scaled.
Let $\sigma_w^T$ be the row vector of weighted column standard deviations of $M_S$ weighted by $w$. 
Scaling $M_S$ to have weighted column variance of 1 is equivalent to the operation $M_S^* = M_S\Sigma$ where $\Sigma = 
\begin{bmatrix}
1/\sigma_{w_1} & 0 & 0\\
0 & \ddots & 0\\
0 & 0 & 1/\sigma_{w_p}
\end{bmatrix}
$. The scaled and centered OLS estimator becomes

\begin{align*}
    \hat{\beta}_{\text{transformed}} &= \Bigg(
\Big((M_S - 1_n\mu_w^T)\Sigma\Big)^T 
W 
\Big((M_S - 1_n\mu_w^T)\Sigma\Big)
\Bigg)^{-1} 
\Big((M_S - 1_n\mu_w^T)\Sigma\Big)^T W y \\
  &= \Big(\Sigma M_S^TWM_S \Sigma - \Sigma M_S^TW1_n\mu_w^T \Sigma - \Sigma \mu_w 1_n^T WM_S \Sigma + \Sigma \mu_w 1_n^T W 1_n \mu_w^T \Sigma \Big)^{-1}    
(\Sigma M_S^TWy - \Sigma \mu_w 1_n^TWy) \\
  &= \Big(M_S^{*^T}WM_S^* - M_S^{*^T}W1_n\mu_w^T \Sigma - \Sigma \mu_w 1_n^T WM_S^* + \Sigma \mu_w 1_n^T W 1_n \mu_w^T \Sigma \Big)^{-1}    
(M_S^{*^T} Wy - \Sigma \mu_w 1_n^TWy)
\end{align*}

\subsection{Covariance of Parameters}

When computing homoskedastic covariances where $var(\epsilon) = k^2 I$, the covariance of $\hat{\beta}_{\text{transformed}}$ can be computed using the standard formula $Cov(\hat{\beta}_{\text{transformed}}) = (\tilde{M}^T \tilde{M})^{-1} \hat{k}^2$ where $\hat{k}^2 = \frac{1}{n - p} \sum_i (y - \hat{y})^2$. For efficient computation, we reuse the expansion

\begin{align*}
(\tilde{M}^T \tilde{M})^{-1} &=
\Bigg(
\Big((M_S - 1_n\mu_w^T)\Sigma\Big)^T 
W 
\Big((M_S - 1_n\mu_w^T)\Sigma\Big)
\Bigg)^{-1} \\
&= \Big(M_S^{*^T}WM_S^* - M_S^{*^T}W1_n\mu_w^T \Sigma - \Sigma \mu_w 1_n^T WM_S^* + \Sigma \mu_w 1_n^T W 1_n \mu_w^T \Sigma \Big)^{-1}.
\end{align*}

Likewise, for heteroskedastic-consistent covariances (\cite{white1980heteroskedasticity}) we expand $(\tilde{M}^T \tilde{M})^{-1} (\tilde{M}^T diag(\epsilon^2) \tilde{M}^T) (\tilde{M}^T \tilde{M})^{-1}$ by replacing $W$ with a diagonal matrix having diagonal entries $w \epsilon^2$.

\section{Predictions on $M_S$}

Fitting OLS on centered and scaled inputs with centers $\mu_w$ and standard deviations $\sigma_w$ yields coefficients that can be used to create predictions for the centered and scaled data, such as $\hat{y} = (M_S - 1_n\mu_w^T)\Sigma \hat{\beta}_{\text{transformed}}$.
However, in practice we will receive new data in the form of $M_S$, not $(M_S - 1_n\mu_w^T)\Sigma$. To predict directly on $M_S$, we compute a $\hat{\beta}_{\text{original}}$ for the original scale of the data, where $\hat{\beta}_{\text{original}} = (I_{p \times p} - 1_p^{(1)} \mu_w^T) \Sigma \hat{\beta}_{\text{transformed}}$, and $1_p^{(1)}$ is a length $p$ column vector with values 1 located at index 1, and 0 everywhere else.

\section{Optimizing Linear Algebra}

There are four key terms throughout the OLS expansions that can be computed efficiently:

\begin{enumerate}
    \item $M_S^{*^T} W M_S^*$.
    This product is symmetric, so only half of it needs to be computed. It can be optimized using sparse matrix multiplications, which are implemented in Eigen (\cite{eigenweb}).
    \item $M_S^{*^T} W1_n\mu_w^T \Sigma$.
    The multiplication can be ordered specifically as $(M_S^{*^T} W1_n) (\mu_w^T \Sigma)$. $(M_S^{*^T} W1_n)$ reduces to weighted column sums of $M_S^{*}$ weighted by $w$. $(\mu_w^T \Sigma)$ reduces to elementwise multiplication between the length $p$ vectors $\mu_w$ and $1 / \sigma_w$.
    \item $\Sigma \mu_w 1_n^T W 1_n \mu_w^T \Sigma$.
    This product is symmetric again. The multiplication can be ordered specifically as $(\Sigma \mu_w) (1_n^T W 1_n) (\mu_w^T \Sigma)$. From 2) $\Sigma \mu_w$ is computed using elementwise multiplication. $(1_n^T W 1_n)$ is the sum of the weights vector, $w$. 
    \item $Wy$ and $1_n^T Wy$. $Wy$ is elementwise multiplication of the length $n$ vectors $w$ and $y$. $1_n^T W y$ sums that elementwise multiplication.
\end{enumerate}

In general, the runtime complexity for estimating OLS is $O(np^2 + p^3)$. When $n > 2p$, it is dominated by constructing $M^T M$, otherwise it is dominated by a pseudoinverse for $(M^T M)^{-1}$. Despite requesting centered inputs, these linear algebra optimizations allow us to form $M_S^{*^T} M_S^*$ using sparse inputs, which combined with simple dense operations can construct $\tilde{M}^T \tilde{M}$ efficiently. However, we cannot invert $\tilde{M}^T \tilde{M}$ efficiently without using dense algebra.

For large $n$, the memory requirements to materialize $\tilde{M}$ can be very large. Using optimized operations, we only materialize $\tilde{M}^T \tilde{M}$, without materializing $\tilde{M}$. The memory requirement to materialize $(M_S, W, \mu_w^T, \sigma_w^T)$, which are the components for the OLS expansions, is $O(np \cdot \text{density} + n)$. Storing $\tilde{M}^T \tilde{M}$ only requires $O(p^2)$ memory.

\subsection{Performance}

In this section we compare the performance of the efficient solver against that of the naive solver, $\hat{\beta}_{\text{transformed}} = (\tilde{M}^T \tilde{M})^{-1} \tilde{M}^T y$, which materializes the dense $\tilde{M}$ matrix. The experiment simulates sparse model matrices with density rates of 0.01, 0.05, 0.1, 0.15, 0.2 and 0.25, with $p = 100$ features. We also vary the sample size from 0.1 million observations to 10 million observations. Results are displayed in the figure below. When data is large and sparse, the efficient solver estimates OLS much faster than the naive solver, up to 35 times as fast. Similarly, the memory used for the efficient solver is $1/\text{density}$ that of the naive solver. 

\begin{figure}[h]
 
\begin{subfigure}{0.5\textwidth}
\includegraphics[width=1\linewidth, height=7cm]{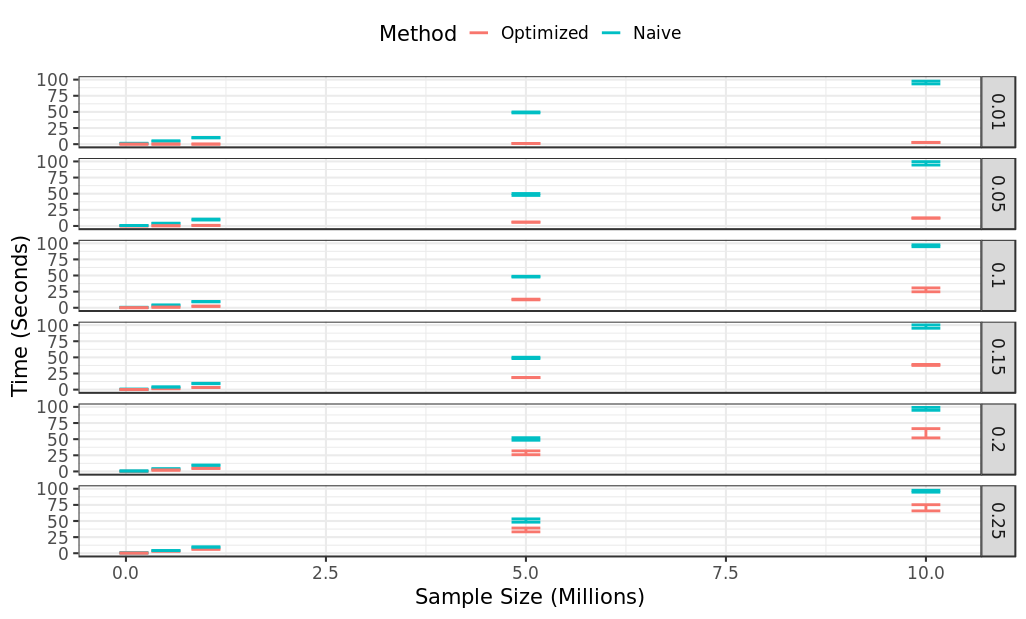}
\caption{Time to estimate centered regression for various sample size and density rates.}
\end{subfigure}
\begin{subfigure}{0.5\textwidth}
\includegraphics[width=1\linewidth, height=7cm]{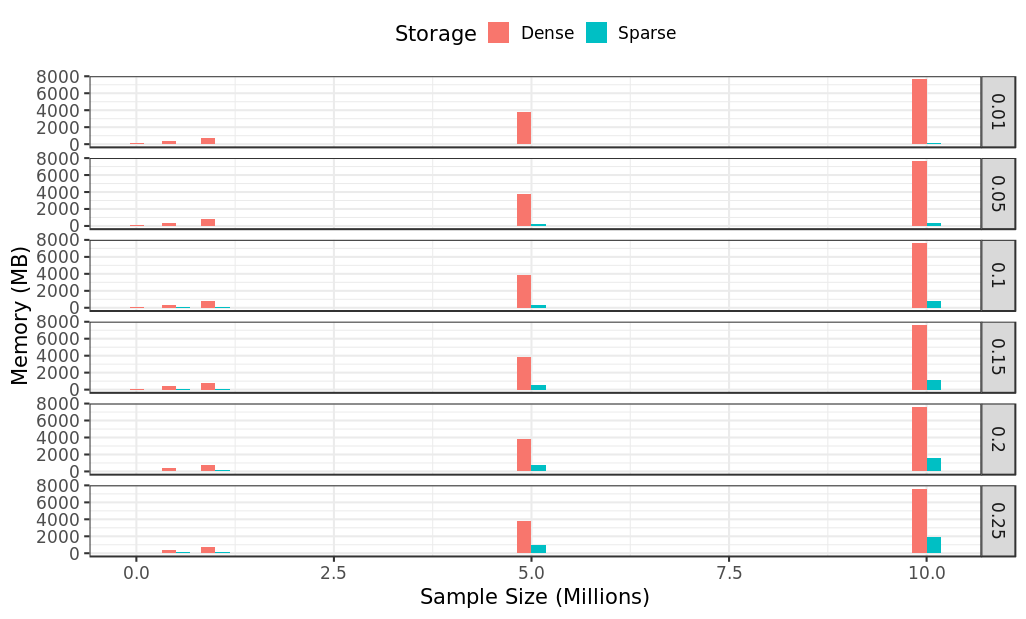}
\caption{Memory usage for various sample size and density rates.}
\end{subfigure}

\end{figure}

\section{Conclusion}

Centered regression has benefits in interpretability and convergence, though centering the features can lead to a computationally expensive regression with dense inputs. We described a computational strategy that expands the standard OLS estimator to take advantage of sparse data structures while still centering the inputs. The resulting implementation resolves the conflict between the desirability of centered regression and the performance benefits of sparse data.

\printbibliography

@book{aiken1991multiple,
  title={Multiple regression: Testing and interpreting interactions},
  author={Aiken, Leona S and West, Stephen G and Reno, Raymond R},
  year={1991},
  publisher={Sage}
}

@article{tibshirani1996regression,
  title={Regression shrinkage and selection via the lasso},
  author={Tibshirani, Robert},
  journal={Journal of the Royal Statistical Society: Series B (Methodological)},
  volume={58},
  number={1},
  pages={267--288},
  year={1996},
  publisher={Wiley Online Library}
}

@article{wong2019efficient,
  title={Efficient Computation of Linear Model Treatment Effects in an Experimentation Platform},
  author={Wong, Jeffrey and Lewis, Randall and Wardrop, Matthew},
  journal={arXiv preprint arXiv:1910.01305},
  year={2019}
}

@inproceedings{seabold2010statsmodels,
  title={statsmodels: Econometric and statistical modeling with python},
  author={Seabold, Skipper and Perktold, Josef},
  booktitle={9th Python in Science Conference},
  year={2010},
}

@article{white1980heteroskedasticity,
  title={A heteroskedasticity-consistent covariance matrix estimator and a direct test for heteroskedasticity},
  author={White, Halbert and others},
  journal={econometrica},
  volume={48},
  number={4},
  pages={817--838},
  year={1980},
  publisher={Princeton}
}

@MISC{eigenweb,
  author = {Ga\"{e}l Guennebaud and Beno\^{i}t Jacob and others},
  title = {Eigen v3},
  howpublished = {http://eigen.tuxfamily.org},
  year = {2010}
 }

\end{document}